\documentstyle{elsart}
\begin{document}
\begin{frontmatter}
\title{Computation of conservation laws for 
nonlinear lattices\thanksref{NSF}}

\author{\"{U}nal G\"{o}kta\c{s}\thanksref{email1}},
\author{Willy Hereman\thanksref{email1}\thanksref{email2}}

\address{Department of Mathematical and Computer Sciences,
Colorado School of Mines, Golden, CO 80401-1887, U.S.A.}

\thanks[NSF]{Research supported in part by NSF
under Grant CCR-9625421.}
\thanks[email1]{E-mail: $\{$ugoktas,whereman$\}$@mines.edu}
\thanks[email2]{Corresponding author.
Phone: (303) 273 3881; Fax: (303) 273 3875.}

\begin{abstract}
An algorithm to compute polynomial conserved densities of 
polynomial nonlinear lattices is presented. 
The algorithm is implemented in {\it Mathematica} and can be used as an 
automated integrability test. 
With the code {\bf diffdens.m}, conserved densities are obtained for 
several well-known lattice equations.
For systems with parameters, the code allows one to determine the 
conditions on these parameters so that a sequence of conservation laws exist.
\end{abstract}


\end{frontmatter}

\section{Introduction}
There are several motives to find the explicit form of conserved densities
of differential-difference equations (DDEs).
The first few conservation laws have a physical meaning, such as 
conservation of momentum and energy. 
Additional ones facilitate the study of both quantitative and 
qualitative properties of solutions \cite{EHandKO}.
Furthermore, the existence of a sequence of conserved densities 
predicts integrability.
Yet, the nonexistence of polynomial conserved quantities does not preclude 
integrability.
Indeed, integrable DDEs could be disguised with a coordinate transformation 
in DDEs that no longer admit conserved densities of polynomial type 
\cite{ASandRY}. 

Another compelling argument relates to the numerical solution of 
partial differential equations (PDEs). In solving integrable
discretizations of PDEs, one should check that conserved quantities 
indeed remain constant.
In particular, the conservation of a positive definite quadratic quantity 
may prevent nonlinear instabilities in the numerical scheme. 
The use of conservation laws in PDE solvers is well-documented 
in the literature \cite{FJH,RJLeV,JMS-S}.

Several methods to test the integrability of DDEs and for solving them 
are available. Solution methods include symmetry reduction 
\cite{DLandPW} and solving the spectral problem \cite{DLandOR} on
the lattice.  
Adaptations of the singularity confinement approach \cite{ARandBGandKT}, 
the Wahlquist-Estabrook method \cite{BD}, and symmetry techniques 
\cite{ICandRY,DLandRY,VSandAS} allow one to investigate integrability. 
The most comprehensive integrability study of nonlinear DDEs was 
done by Yamilov and co-workers (see e.g. \cite{ASandRY,RY}).
Their papers provide a classification of semi-discrete 
equations possessing infinitely many local conservation laws. 
Using the formal symmetry approach, they derive the necessary and sufficient 
conditions for the existence of local conservation laws, and provide
an algorithm to construct them. 

In \cite{UGandWHa,UGandWHb}, we gave an algorithm to compute
conserved densities for systems of nonlinear evolution equations.
The algorithm is based on the concept of invariance of equations under 
dilation symmetries. Therefore, it is inherently limited to polynomial 
densities of polynomial systems.  
Recently an extension of the algorithm towards DDEs was outlined
in \cite{UGandWHandGE}.
Here we present a full description of our algorithm, which can 
be implemented in any computer algebra language. 
We also provide details about our software package 
{\bf diffdens.m} \cite{UGandWHc} in {\it Mathematica\/} \cite{SW},
which automates the tedious computation of closed-form expressions
for conserved densities.

Following basic definitions in Section 2, the algorithm
is given in Section 3.
To illustrate the method we use the Toda lattice \cite{MT}, 
a parameterized Toda lattice \cite{ARandBGandKT}, and the discretized
nonlinear Schr\"odinger (NLS) equation \cite{MAandJLa,MAandJLb,MAandBH}. 
In Section 4, we list results for an extended Lotka-Volterra
equation \cite{XBHandRKB}, a discretized modified KdV equation \cite{MAandPC},
a network equation \cite{MAandPC}, and some other lattices \cite{ASandRY}.
The features, scope and limitations of our code {\bf diffdens.m} are described 
in Section 5, together with instructions for the user. 
In Section 6, we draw some conclusions.
\section{Definitions}

\subsection{Conservation laws}

Consider a system of DDEs which is continuous in time, and discretized 
in the single space variable, 
\begin{equation}\label{multisys}
{\dot{\bf u}}_n = 
{\bf F} (...,{\bf u}_{n-1}, {\bf u}_{n}, {\bf u}_{n+1},...), 
\end{equation}
where ${\bf u}_{n}$ and ${\bf F}$ are vector dynamical variables with 
any number of components. 
For simplicity of notation, the components of ${\bf u}_n $ will be denoted 
by $u_n, v_n,$ etc.
We assume that ${\bf F}$ is polynomial with constant coefficients. 
If DDEs are non-polynomial or of higher order in $t$, we assume that they
can be recast in the form (\ref{multisys}).

A local {\em conservation law\/} is defined by
\begin{equation}\label{conslaw}
{\dot{\rho}}_n = J_n - J_{n+1},
\end{equation}
which is satisfied on all solutions of (\ref{multisys}).
The function $\rho_n $ is the {\em conserved density} and $J_n$ is the 
associated {\em flux.} Both are assumed to be polynomials 
in ${\bf u}_n$ and its shifts. 

Obviously,
$
{\mathrm{d} \over \mathrm{dt}} ( \sum_{n} \rho_n ) = 
   \sum_{n} {\dot{\rho}}_n = \sum_{n} (J_n - J_{n+1}),  
$ 
and this telescopic series vanishes for a bounded periodic lattice
or a bounded lattice resting at infinity. 
In that case, $ \sum_{n} \rho_n $ is constant in time. 
So, we have a conserved quantity.

\begin{exmp}
{\rm
Consider the one-dimensional Toda lattice 
\cite{MT}
\begin{equation}\label{orgtoda}
{\ddot{y}}_n = \exp{(y_{n-1} - y_n)} - \exp{(y_n - y_{n+1})},
\end{equation}
where $y_n$ is the displacement from equilibrium of the 
$n\/$th unit mass under an exponential decaying interaction 
force between nearest neighbors. 

With the change of variables, 
\begin{equation} \label{todamotion}
u_n = {\dot{y}}_n, \quad\quad\quad  v_n = \exp{(y_{n} - y_{n+1})}, 
\end{equation}
lattice (\ref{orgtoda}) can be written in polynomial form
\begin{equation}\label{todalatt}
{\dot{u}}_n = v_{n-1} - v_n, \quad\;\;\; {\dot{v}}_n = v_n (u_n - u_{n+1}).
\end{equation}
System (\ref{todalatt}) is completely integrable \cite{MT,MH}. 
The first two density-flux pairs 
\begin{equation} \label{todapair}
\rho_n^{(1)} = u_n, \quad J_n^{(1)} = v_{n-1}, \;\;\;{\rm and}\;\;\;
\rho_n^{(2)} = {\textstyle{1 \over 2}} u_n^2 + v_n, 
\quad J_n^{(2)} = u_n v_{n-1},
\end{equation}
can be easily be computed by hand. 
}
\end{exmp}

\subsection{Equivalence criterion}

We introduce a few concepts that will be used in our algorithm.
Let $D$ denote the {\it shift-down} operator and $U$ the 
{\it shift-up} operator. Both are defined on the set of all monomials
in ${\bf u}_n$ and its shifts.
If $m$ is such a monomial then $D m = m |_{n \rightarrow n-1} $ 
and $U m = m |_{n \rightarrow n+1}. $ 
For example, $ D u_{n+2} v_{n} = u_{n+1} v_{n-1} $ and 
$ U u_{n-2} v_{n-1} = u_{n-1} v_{n}. $ 
It is easy to verify that compositions of $D$ and $U$ define an 
{\em equivalence relation\/} on monomials.
Simply stated, all shifted monomials are {\it equivalent}, e.g. 
$u_{n-1} v_{n+1} \, \equiv \, u_{n+2} v_{n+4} \, \equiv \, u_{n-3} v_{n-1}.$
This equivalence relation holds for any function of the dependent variables,
but for the construction of conserved densities we will apply it only
to monomials.

In the algorithm we will use the following 
{\it equivalence criterion}: if two monomials
$m_1$ and $m_2$ are equivalent, $m_1 \, \equiv m_2,$ then
$m_1 \,=\, m_2 + [M_n - M_{n+1} ] $ for some polynomial $M_n$ 
that depends on ${\bf u}_n$ and its shifts.
For example, $ u_{n-2} u_n \, \equiv \, u_{n-1} u_{n+1} $
since $ u_{n-2} u_n = u_{n-1} u_{n+1} + [u_{n-2} u_{n} - u_{n-1} u_{n+1}] 
= u_{n-1} u_{n+1} + [M_n - M_{n+1}], $ with $M_n = u_{n-2} u_{n}.$

The {\it main} representative of an equivalence class is
the monomial of that class with $n$ as {\it lowest\/} label 
on $u$ (or $v$). 
For example, $u_n u_{n+2}$ is the main representative of the class 
with elements $u_{n-1} u_{n+1}, u_{n+1} u_{n+3},$ etc. 
Lexicographical ordering is used to resolve conflicts. For example, 
$u_n v_{n+2} $ (not $u_{n-2} v_n$) is the main representative in 
the class with elements $u_{n-3} v_{n-1}, u_{n+2} v_{n+4},$ etc.

\section{Algorithm}

Scaling invariance, which results from a special Lie-point symmetry, 
is an intrinsic property of many integrable nonlinear PDEs and DDEs. 
Indeed, observe that (\ref{todalatt}), and the couples 
$\rho_n^{(1)}, J_n^{(1)}$ and $\rho_n^{(2)}, J_n^{(2)}$ 
in (\ref{todapair}) 
(after inserting them in (\ref{conslaw})), 
are invariant under the dilation (scaling) symmetry
\begin{equation}\label{symtoda}
(t, u_n, v_n) \rightarrow (\lambda t, \lambda^{-1} u_n, {\lambda}^{-2} v_n),
\end{equation}
where $\lambda$ is an arbitrary parameter. Stated differently, 
$u_n$ corresponds to one derivative with respect to $t;$ denoted by
$ u_n \sim \frac{\mathrm{d}}{\mathrm{dt}}.$
Similarly, $v_n \sim \frac{{\mathrm{d}}^2}{{\mathrm{dt}}^2}.$

Our three-step algorithm exploits the symmetry (\ref{symtoda}) 
to find conserved densities.
\vskip 3pt
\noindent
{\em Step 1: Determine the weights of variables}

The {\em weight\/}, $w,$ of a variable is by definition equal to the number 
of derivatives with respect to $t\/$ the variable carries. 
Weights are non-negative, rational, and independent of $n.$
Without loss of generality we set $w({\mathrm{d} \over \mathrm{dt}}) = 1.$

The {\em rank} of a monomial is defined as the total weight of the monomial.
For now we assume that all the terms (monomials) in a particular equation 
have the same rank. This property is called {\em uniformity in rank}. The
uniformity in rank condition leads to a system of equations for the unknown
weights.
Different equations in the vector equation (\ref{multisys}) may have 
different ranks.

\vfill
\newpage
\begin{exmp}
{\rm
Requiring uniformity in rank for each equation in
(\ref{todalatt}) allows one to compute the weights of the dependent
variables. Indeed,
\begin{equation} \label{uniformity}
w(u_n) + 1 = w(v_n), \quad w(v_n) + 1 = w(u_n) + w(v_n), 
\end{equation} 
yields
$ w(u_n) = 1, \, w(v_n) = 2, $ which is consistent with (\ref{symtoda}).
}
\end{exmp}

\vskip 3pt
\noindent
{\em Step 2: Construct the form of the density}

This step involves finding the building blocks (monomials) of a 
polynomial density with prescribed rank $R$. 
All terms in the density must have the same
rank $R$. Since we may introduce parameters with weights 
(see {\rm Example 6}), 
the fact that the density will be a sum of monomials of uniform rank does 
not necessarily imply that the density must be uniform in rank with 
respect to the dependent variables.

Let $\cal V$ be the list of all the variables with positive weights, including
parameters with weight. The following procedure is used to determine
the form of the density of rank $R$:
\begin{itemize}
\item Form the set $\cal G$ of all monomials of rank $R$ or less by 
taking all appropriate combinations of different powers of the variables 
in $\cal V$.
\item For each monomial in $\cal G$, introduce the appropriate number of
derivatives with respect to $t$ so that all the monomials exactly have
weight $R$. Gather in set $\cal H$ all the terms that result from
computing the various derivatives.
\item Identify the monomials that belong to the same equivalence classes
and replace them by the main representatives. 
Call the resulting simplified set $\cal I$, which consists of the 
{\it building blocks\/} of the density with desired rank $R$.
\item Linear combination of the elements in ${\cal I}$ with
constant coefficients $c_i$ gives the form of polynomial density of rank $R.$
\end{itemize}

\begin{exmp}
{\rm 
Continuing with (\ref{todalatt}),
we compute the form of the density of rank $R=3.$ From ${\cal V} 
= \{ u_n, v_n \}$ we build 
$ {\cal G} = \{ u_n^3, u_n^2, u_n v_n, {u_n}, v_n \}. $ 
Next, introduce $t$-derivatives,
so that each term exactly has rank $3.$ Thus, using (\ref{todalatt}), 
\begin{eqnarray*}
&& \!\!\!\!\! 
{{\mathrm{d}}^0 \over {\mathrm{dt}}^0} ( u_n^3 )
   = u_n^3 , \;\;\;\;\quad 
{{\mathrm{d}}^0 \over {\mathrm{dt}}^0} ( u_n v_n )
   = u_n v_n , \\
&& \!\!\!\!\!  \\
&& \!\!\!\!\! 
{{\mathrm{d}} \over {\mathrm{dt}}} ( {u_n}^2 )
   = 2 u_n {\dot{u}}_n = 2 u_n v_{n-1} - 2 u_n v_n ,  \;\;\;\;\quad
{{\mathrm{d}} \over {\mathrm{dt}}} ( v_n )
   = {\dot{v}}_n =  u_n v_n -  u_{n+1} v_n, \\
&&  \!\!\!\!\!  \\
&& 
\!\!\!\!\! 
{{\mathrm{d}}^2 \over {\mathrm{dt}}^2} ( u_n ) 
   = {{\mathrm{d}} \over {\mathrm{dt}}} ( {\dot{u}}_n ) 
   = {{\mathrm{d}} \over {\mathrm{dt}}} ( v_{n-1} - v_n ) 
   = u_{n-1} v_{n-1} - u_{n} v_{n-1} - u_n v_n + u_{n+1} v_n .
\end{eqnarray*}
Gather the resulting terms in set
$
{\cal H}  = 
\{ u_n^3, u_n v_{n-1} , u_n v_n , u_{n-1} v_{n-1} , u_{n+1} v_n \} . 
$
Since $ u_{n+1} v_n \equiv u_n v_{n-1} \,$ and
$ u_{n-1} v_{n-1} \equiv u_n v_n $ in $\cal H$, 
$ u_{n+1} v_n $ is replaced by $u_n v_{n-1},$ and $ u_{n-1} v_{n-1} $
by $ u_n v_n.$ 
Hence, we obtain 
$ {\cal I} = \{ u_n^3 , u_n v_{n-1} , u_n v_n \}. $
\vskip 1pt
\noindent
Linear combination of the monomials in ${\cal I}$ 
gives the form of the density:
\begin{equation}\label{formrho3toda}
\rho_n = c_1 \, u_n^3 + c_2 \, u_n v_{n-1} + c_3 \, u_n v_n . 
\end{equation}
}
\end{exmp}
\vskip 3pt
\noindent
{\em Step 3: Determine the unknown coefficients in the density}

The following procedure simultaneously determines the constants $c_i$ 
and the form of the flux $J_n$: 
\begin{itemize}
\item Compute ${\dot{\rho}}_n$ and use (\ref{multisys}) to remove
all the $t\/$-derivatives. 
\item Regarding (\ref{conslaw}), the resulting expression must match 
the pattern $J_n - J_{n+1}$. Use the equivalence criterion to modify
${\dot{\rho}}_n$. The goal is to introduce the main representatives
and to identify the terms that match the pattern.
\item The part that does not match the pattern must
vanish identically for any combination of the components of ${\bf u}_n$ 
and their shifts. This leads to a {\it linear} system $\cal S$ in 
the unknowns $c_i$. 
If $\cal S$ has parameters, careful analysis leads to conditions
on these parameters guaranteeing the existence of densities. 
See \cite{UGandWHa} for a description of this compatibility analysis.
\item The flux is the first piece in the pattern $[ J_n - J_{n+1} ].$
\end{itemize}

\begin{exmp}
{\rm
Carrying on with (\ref{todalatt}), we determine the 
coefficients $c_1$ through $c_3$ in (\ref{formrho3toda}) by requiring 
that (\ref{conslaw}) holds. Simultaneously, we determine $J_n.$

Compute ${\dot{\rho}}_n$ using (\ref{formrho3toda}). Use 
(\ref{todalatt}) to remove ${\dot{u}_n}, {\dot{v}_n}, $ etc. 
After grouping the terms
\begin{eqnarray}
{\dot \rho}_n 
& \,= \,& ( 3 c_1 - c_2 ) u_n^2 v_{n-1} + (c_3 - 3 c_1 ) u_n^2 v_n
          + (c_3 - c_2) v_{n-1} v_n  \nonumber \\
& &       + c_2 u_{n-1} u_n v_{n-1} + c_2 v_{n-1}^2
          - c_3 u_n u_{n+1} v_n - c_3 v_n^2. 
\end{eqnarray}
Use the equivalence criterion to modify ${\dot \rho}_n$.
For instance, replace $u_{n-1} u_n v_{n-1}$ by
$u_{n} u_{n+1} v_n + [u_{n-1} u_n v_{n-1} - u_{n} u_{n+1} v_{n}]$. 
In terms of main representatives, 
\begin{eqnarray}
{\dot \rho}_n 
& \,= \,& ( 3 c_1 - c_2 ) u_n^2 v_{n-1} + (c_3 - 3 c_1 ) u_n^2 v_n 
\nonumber \\
& &+ (c_3 - c_2) v_{n} v_{n+1} + 
[(c_3 - c_2) v_{n-1} v_n -(c_3-c_2) v_{n} v_{n+1}] \nonumber \\
& & + c_2 u_{n} u_{n+1} v_{n} + 
    [c_2 u_{n-1} u_n v_{n-1} - c_2 u_{n} u_{n+1} v_{n}]  \nonumber \\
& & + c_2 v_{n}^2 + [c_2 v_{n-1}^2 - c_2 v_{n}^2 ] 
- c_3 u_n u_{n+1} v_n - c_3 v_n^2. 
\end{eqnarray}
Next, group the terms outside of the square brackets and move the pairs
inside the square brackets to the bottom. Rearrange the latter terms 
so that they match the pattern $[J_n - J_{n+1}]$. Hence,
\begin{eqnarray}
{\dot \rho}_n 
& \,= \,& ( 3 c_1 - c_2 ) u_n^2 v_{n-1} + (c_3 - 3 c_1 ) u_n^2 v_n 
\nonumber \\
& & + (c_3 - c_2) v_{n} v_{n+1} + (c_2 - c_3) u_{n} u_{n+1} v_{n} 
+ ( c_2 - c_3 ) v_n^2 \nonumber \\
& & + [ 
\{ 
(c_3 - c_2) v_{n-1} v_n + c_2 u_{n-1} u_n v_{n-1} + c_2 v_{n-1}^2  
\} \nonumber \\
& & - \{ 
(c_3-c_2) v_{n} v_{n+1} + c_2 u_{n} u_{n+1} v_{n} + c_2 v_{n}^2 
\} ].
\end{eqnarray}
The first piece inside the square brackets determines
\begin{equation} \label{formflux3toda}
J_n = (c_3 - c_2) v_{n-1} v_n + c_2 u_{n-1} u_n v_{n-1} + c_2 v_{n-1}^2. 
\end{equation}
The terms outside the square brackets must all vanish piece by piece, 
yielding 
\begin{equation}\label{todasystem}
{\cal S} = \{ 3 c_1 - c_2 = 0, c_3 - 3 c_1 = 0, c_2 - c_3 = 0 \}. 
\end{equation}
The solution is $ 3 c_1 = c_2 = c_3.$ 
Since densities can only be determined up to a multiplicative constant, 
we choose $ c_1 = {\textstyle \frac{1}{3}}, \, c_2 = c_3 = 1,$ 
and substitute this into (\ref{formrho3toda}) and (\ref{formflux3toda}). 
The explicit forms of the density and the flux follow:
\begin{equation} \label{rhofortoda}
\rho_n = {\textstyle{1 \over 3}} \, u_n^3 + u_n ( v_{n-1} + v_n ),
\quad\;\;\; J_n = u_{n-1} u_n v_{n-1} + v_{n-1}^2.
\end{equation}
}
\end{exmp}

\begin{exmp}
{\rm
To illustrate how the algorithm works for DDEs with parameters, consider
the parameterized Toda lattice 
\begin{equation}\label{partodalatt}
{\dot{u}}_n = \alpha \; v_{n-1} - v_n, \quad \quad
{\dot{v}}_n = v_n \; (\beta \; u_n - u_{n+1}),
\end{equation} 
where $\alpha$ and $\beta$ are {\it nonzero} parameters without weight.
In \cite{ARandBGandKT} it was shown that (\ref{partodalatt})
is completely integrable if and only if $\alpha = \beta = 1$; but then 
(\ref{partodalatt}) is (\ref{todalatt}).

Using our algorithm, one can easily compute the {\it compatibility conditions}
for $\alpha$ and $\beta$, so that (\ref{partodalatt}) admits a polynomial 
conserved densities of, say, rank 3.
The steps are the same as for (\ref{todalatt}). 
However, (\ref{todasystem}) must be replaced by 
\[
\!\!\!\!\!\!\!\!\!\!
{\cal S} = \{ 3 \alpha c_1 - c_2 = 0, \beta c_3 - 3 c_1 = 0, 
\alpha c_3 - c_2 = 0, \beta c_2 - c_3 = 0, \alpha c_2 - c_3 = 0 \}. 
\]
A non-trivial solution $ 3 c_1 = c_2 = c_3 $ will exist if and only if
$ \alpha = \beta = 1.$

Analogously, (\ref{partodalatt}) has density $\rho_n^{(1)} = u_n$ of rank 1 
if $ \alpha = 1, $ and density
$ \rho_n^{(2)} = {\textstyle \frac{\beta}{2}} u_n^2 + v_n $ of rank 2 if 
$ \alpha \, \beta = 1.$ 
Only when $ \alpha = \beta = 1 $ will (\ref{partodalatt}) have
conserved densities of rank $\ge$ 3: 
\begin{eqnarray}
\rho_n^{(3)} &=& {\textstyle{1 \over 3}} u_n^3 + u_n (v_{n-1} + v_n), \\
\rho_n^{(4)} &=& {\textstyle{1 \over 4}} u_n^4 + u_n^2 (v_{n-1} + v_n)
+ u_n u_{n+1} v_n + {\textstyle{1 \over 2}} v_n^2 + v_n v_{n+1}, \\
\rho_n^{(5)} &=& {\textstyle{1 \over 5}} u_n^5 + u_n^3 (v_{n-1} + v_n ) 
+ u_n u_{n+1} v_n ( u_n + u_{n+1}) \nonumber \\
&& + u_n v_{n-1} (v_{n-2} + v_{n-1} + v_n )
+ u_n  v_n ( v_{n-1} + v_n +  v_{n+1} ).
\end{eqnarray}
Ignoring irrelevant shifts in $n,$ these densities agree with the 
results in \cite{MH}. 
}
\end{exmp}

\begin{exmp}
{\rm
In \cite{MAandJLa,MAandJLb}, Ablowitz and Ladik studied the 
following integrable discretization of the NLS equation:
\begin{equation} \label{orgabllad}
i \, {\dot{u}}_n =
u_{n+1}- 2 u_n + u_{n-1} +  u_n^{*} u_n (u_{n+1} + u_{n-1}), 
\end{equation}
where $u_n^{*}$ is the complex conjugate of $u_n.$
Instead of splitting $u_n$ into its real and imaginary parts, we treat
$u_n$ and $v_n = u_n^{*}$ as independent variables and 
augment (\ref{orgabllad}) with its complex conjugate. 
Absorbing $i$ in the scale on $t,$
\begin{eqnarray} \label{abllad}
{\dot{u}}_n 
&=& u_{n+1} - 2 u_n + u_{n-1} + u_n v_n (u_{n+1} + u_{n-1}), 
\nonumber \\ 
{\dot{v}}_n 
&=& -( v_{n+1} - 2 v_n + v_{n-1} ) -  u_n v_n (v_{n+1} + v_{n-1}).
\end{eqnarray}
Since $v_n = u_n^{*} $ we have $w(v_n) = w(u_n).$ 
Neither of the equations in (\ref{abllad}) is uniform in rank. 
To circumvent this problem we introduce an auxiliary parameter 
$\alpha$ with weight, and replace (\ref{abllad}) by
\begin{eqnarray} \label{ablladnew}
{\dot{u}}_n 
&=& \alpha ( u_{n+1} - 2 u_n + u_{n-1} ) + u_n v_n (u_{n+1} + u_{n-1}), 
\nonumber \\
{\dot{v}}_n 
&=& - \alpha ( v_{n+1} - 2 v_n + v_{n-1} ) -  u_n v_n (v_{n+1} + v_{n-1}).
\end{eqnarray}
This extra freedom allows us to impose uniformity in rank:
\begin{eqnarray}
w(u_n) + 1 &=& w(\alpha) + w(u_n) = 2 w(u_n) + w(v_n) = 3 w(u_n), \\
w(v_n) + 1 &=& w(\alpha) + w(v_n) = 2 w(v_n) + w(u_n) = 3 w(v_n), 
\end{eqnarray}
which yields $ w(u_n) = w(v_n) = {\textstyle \frac{1}{2}}, w(\alpha) = 1, $
or, $ u_n^2 \sim v_n^2 \sim \alpha \sim {\mathrm{d} \over \mathrm{dt}}.$

We show how to get the building blocks of the density 
of rank ${\textstyle \frac{3}{2}}$. 
In this case ${\cal V} = \{ u_n, v_n, \alpha \}$ 
and  
$
{\cal G} = \{ u_n, v_n, \alpha, u_n^2, u_n v_n, v_n^2,
              \alpha u_n, u_n^3, \alpha v_n, u_n^2 v_n, 
              u_n v_n^2, v_n^3 \}.
$
The monomials $\alpha u_n, u_n^3, \alpha v_n, u_n^2 v_n, u_n v_n^2$
and $v_n^3$ have already rank ${\textstyle \frac{3}{2}}$, 
so no derivatives are needed.
The monomials $u_n$ and $v_n$ will have rank ${\textstyle \frac{3}{2}}$ 
after introducing 
${\mathrm{d} \over \mathrm{dt}}$. There is no way for the remaining 
monomials $\alpha, {u_n}^2 , u_n v_n$ and ${v_n}^2$ to have rank 
${\textstyle \frac{3}{2}}$ after differentiation with respect to $t$. 
Therefore, they are rejected. 
The remaining intermediate steps lead to
\[
\!\!\!\!\!\!\!\!\!
{\cal I} = \{ \alpha u_n, u_n^3, \alpha v_n, u_n^2 v_n,
   u_n u_{n+1} v_n, u_n v_{n-1} v_n, u_n v_n^2,
   v_n^3, u_n u_{n+1} v_{n+1}, u_n v_n v_{n+1} \}.
\]
Although uniformity in rank is essential for the first two steps 
of the algorithm, after the second step, we may already set $\alpha = 1.$
The computations now proceed as in the previous example. 
The trick of introducing one or more extra parameters
with weights can always be attempted if any equation in (\ref{multisys})
lacks uniformity in rank.

We list some conserved densities of (\ref{abllad}): 
\begin{eqnarray}
\!\!\!\!\!\!\!\!\!\!
\rho_n^{(1)} &=& c_1 u_n v_{n-1} + c_2 u_n v_{n+1}, \\ 
\!\!\!\!\!\!\!\!\!\!
\rho_n^{(2)} &=& c_1 
(
{\textstyle{1 \over 2}} u_n^2 v_{n-1}^2 + 
u_n u_{n+1} v_{n-1} v_n + u_n v_{n-2} ) \nonumber \\
\!\!\!\!\!\!\!\!\!\!
&+& c_2 (
{\textstyle{1 \over 2}} u_n^2 v_{n+1}^2 + 
u_n u_{n+1} v_{n+1} v_{n+2} + u_n v_{n+2}), \\
\!\!\!\!\!\!\!\!\!\!
\rho_n^{(3)} &=& c_1 [
{\textstyle{1 \over 3}} u_n^3 v_{n-1}^3 + 
u_n u_{n+1} v_{n-1} v_n ( u_n v_{n-1} + u_{n+1} v_n + u_{n+2} v_{n+1} ) 
\nonumber \\
\!\!\!\!\!\!\!\!\!\!\!\!
&+& u_n v_{n-1} (u_n v_{n-2} + u_{n+1} v_{n-1} )
+ u_n v_n (u_{n+1} v_{n-2} + u_{n+2} v_{n-1} ) + u_n v_{n-3} ] 
\nonumber \\
\!\!\!\!\!\!\!\!\!\!\!\!\!
&+&  c_2 [ 
{\textstyle{1 \over 3}} u_n^3 v_{n+1}^3 + 
u_n u_{n+1} v_{n+1} v_{n+2} (u_n v_{n+1} +u_{n+1} v_{n+2} + u_{n+2} v_{n+3}) 
\nonumber \\
\!\!\!\!\!\!\!\!\!\!
&+&\! u_n v_{n+2} (u_n v_{n+1} \!+\! u_{n+1} v_{n+2} )
\!+\! u_n v_{n+3} (u_{n+1} v_{n+1} \!+\! u_{n+2} v_{n+2} ) \!+\! u_n v_{n+3}].
\end{eqnarray}
Our results confirm those in \cite{MAandJLa}.
Also, if defined on an infinite interval, (\ref{orgabllad}) admits 
infinitely many independent conserved densities \cite{MAandJLa}.
Although it is a constant of motion, we cannot find the Hamiltonian of 
(\ref{orgabllad}), 
\begin{equation} \label{hamiltonian}
H = -i \sum_n 
[ u_n^{*} ( u_{n-1} + u_{n+1} ) - 2 \ln (1 + u_n u_n^{*}) ], 
\end{equation} 
for it has a logarithmic term \cite{MAandBH}. 
}
\end{exmp}

\section{More examples}

\subsection{An extended Lotka-Volterra equation}

In \cite{XBHandRKB}, Hu and Bullough considered an extended version of 
the Lotka-Volterra equation: 
\begin{equation}\label{extvolterra}
{\dot{u}}_n = \sum_{r=1}^{k-1} (u_{n-r} - u_{n+r}) u_n. 
\end{equation}
For $k=2,$ (\ref{extvolterra}) is the well-known Lotka-Volterra
equation, for which the densities were presented in \cite{UGandWHandGE}.

We computed five densities of (\ref{extvolterra}) for the case $k = 3$.
The first three are: 
\begin{eqnarray}
\rho_1 &=& u_n, \quad\quad\quad\quad\quad
\rho_2 = \frac{1}{2} u_n^2 + u_n ( u_{n+1} + u_{n+2} ), \\
\rho_3 &=& \frac{1}{3} u_n^3 + u_n^2 ( u_{n+1} + u_{n+2} ) 
+ u_n ( u_{n+1}^2 + u_{n+2}^2 ) \nonumber \\
& & + u_n u_{n+1} ( u_{n+2} + u_{n+3} ) 
+ u_n u_{n+2} ( u_{n+1} + u_{n+3} + u_{n+4} ) . 
\end{eqnarray}

For $k = 4$, we also computed five densities of (\ref{extvolterra}). 
The first three are:
\begin{eqnarray}
\!\!\!\!\!\!\!\!\!
\rho_1 &=& u_n, \quad\quad\quad\quad\quad
\rho_2 = \frac{1}{2} u_n^2 + u_n ( u_{n+1} + u_{n+2} + u_{n+3} ),
\\
\!\!\!\!\!\!\!\!\!
\rho_3 &=& \frac{1}{3} u_n^3 + u_n^2 ( u_{n+1} + u_{n+2} + u_{n+3} )
+ u_n (u_{n+1} +u_{n+2} + u_{n+3})^2 \nonumber \\
\!\!\!\!\!\!\!\!\!
&+&  u_n u_{n+3} ( u_{n+4} + u_{n+5} + u_{n+6} )
+ u_n u_{n+4} ( u_{n+1} + u_{n+2} ) + u_n u_{n+2} u_{n+5} . 
\end{eqnarray}

We computed four densities of (\ref{extvolterra}) for $k=5.$ 
To save space we do not list them here. The integrability and other 
properties of (\ref{extvolterra}) are discussed in \cite{XBHandRKB}.

\subsection{A discretized modified KdV equation}

In \cite{MAandPC}, we found the following integrable discretization of the 
MKdV equation:
\begin{equation}\label{discmkdv}
{\dot{u}}_n = (1 + u_n^2) (u_{n+1} - u_{n-1}).
\end{equation}
We computed four densities of (\ref{discmkdv}). The first three are:
\begin{eqnarray}
\rho_n^{(1)} &=& u_n {u_{n+1}}, \quad \quad \quad 
\rho_n^{(2)} = {\textstyle \frac{1}{2}} u_n^2 u_{n+1}^2 
+ u_n {u_{n+2}} ( 1 + u_{n+1}^2 ), \\
\rho_n^{(3)} &=& {\textstyle \frac{1}{3}} u_n^3 u_{n+1}^3 
+ {u_n} u_{n+1} {u_{n+2}} ( u_n + u_{n+2} ) ( 1 + u_{n+1}^2 ) 
\nonumber \\ & & 
+ u_n u_{n+3} ( 1 + u_{n+1}^2 ) ( 1 + u_{n+2}^2 ). 
\end{eqnarray}

\subsection{Self-dual network equations}

The integrable nonlinear self-dual network equations 
\cite{ASandRY,MAandPC} can be written as:
\begin{equation}\label{dual}
{\dot{u}}_n = (1 + u_n^2) (v_{n} - v_{n-1}), \quad \quad \quad
{\dot{v}}_n = (1 + v_n^2) (u_{n+1} - u_n).
\end{equation}
We computed the first four densities of (\ref{dual}).
The first three are
\begin{eqnarray}
\rho_n^{(1)} &=& u_n v_{n-1} + u_n v_n, \\
\rho_n^{(2)} &=& {\textstyle \frac{1}{2}} u_n^2 ( v_{n-1}^2 + v_n^2 )
+ u_n {u_{n+1}} ( 1 + v_n^2 ) + v_n ( u_n^2 v_{n-1} + v_{n+1} ), \\
\rho_n^{(3)} &=& {\textstyle \frac{1}{3}} u_n^3 ( v_{n-1}^3 + v_n^3 )
+ u_n u_{n+1} ( u_n v_{n-1} + u_{n+1} v_n + u_n v_n) ( 1 + v_n^2) 
\nonumber \\
& & + u_n v_{n-2} ( 1 + v_{n-1}^2 ) 
+ u_n v_{n-1} v_n ( v_{n-1} + v_n ) ( 1 + u_n^2 ) \nonumber \\
& & + u_n v_{n+1} ( 1+ u_{n+1}^2 ) ( 1 + v_n^2 ).
\end{eqnarray}

\subsection{Generalized lattices} 

Shabat and Yamilov \cite{ASandRY} studied the following integrable 
Volterra lattice:
\begin{equation}\label{ravilvolterra}
{\dot{u}}_n = u_n (v_{n+1} - v_{n}), \quad\quad\quad
{\dot{v}}_n = v_n (u_{n} - u_{n-1}).
\end{equation}
With our program we computed the first four densities for this system:
\begin{eqnarray}
\!\!\!\!\!\!\!\!\!\!\!\!\!\!\!\!
\rho_n^{(1)} &=& u_n + v_n, \quad\quad\quad\quad\quad\quad\quad
\rho_n^{(2)} = 
{\textstyle{1 \over 2}} ( u_n^2 + v_n^2 ) + u_n (v_{n} + v_{n+1}), \\
\!\!\!\!\!\!\!\!\!\!\!\!\!\!\!\!
\rho_n^{(3)} &=& 
{\textstyle{1 \over 3}} ( u_n^3 \!+\! v_n^3 ) 
\!+\! u_n^2 ( v_{n} \!+\! v_{n+1}) \!+\! u_n ( v_{n}^2 \!+\!  v_{n+1}^2 )
\!+\! u_n v_{n+1} (u_{n+1} \!+\! v_{n}), \\
\rho_n^{(4)} &=&
{\textstyle{1 \over 4}} ( u_n^4 + v_n^4 )
+ u_n^3 ( v_{n} + v_{n+1}) 
+ {\textstyle{3 \over 2}} u_n^2 ( v_{n}^2  +  v_{n+1}^2 ) \nonumber \\
& & + u_n ( v_{n}^3  +  v_{n+1}^3 )
+ 2 u_n v_{n+1} ( u_n + u_{n+1} ) \nonumber \\
& & + u_n u_{n+1} v_{n+1} ( u_n + u_{n+1} + v_n + v_{n+2} )
+ u_n v_n v_{n+1} ( v_n + v_{n+1} ). 
\end{eqnarray}

In \cite{ASandRY}, the following Hamiltonian lattice is also listed:
\begin{equation}\label{ravilsys}
{\dot{u}}_n = u_{n+1} + u_n^2 v_n, \quad\quad
{\dot{v}}_n = -v_{n-1} - u_n v_n^2 .
\end{equation}
Four densities of (\ref{ravilsys}) are:
\begin{eqnarray}
\!\!\!\!\!\!\!\!\!\!\!\!\!\!\!\!
\rho_n^{(1)} &=& u_n v_n, \quad\quad\quad\quad\quad
\rho_n^{(2)} = 
{\textstyle{1 \over 2}} u_n^2 v_n^2 + u_n v_{n-1} , \\
\!\!\!\!\!\!\!\!\!\!\!\!\!\!\!\!
\rho_n^{(3)} &=& 
{\textstyle{1 \over 3}} u_n^3 v_n^3  
+ {u_n} {v_n} ( u_n v_{n-1} + u_{n+1} v_n )
+ u_n v_{n-2} , \\
\rho_n^{(4)} &=&
{\textstyle{1 \over 4}} u_n^4 v_n^4
+ u_n^2 v_n^2 ( u_n v_{n-1} + u_{n+1} v_{n}) 
+ {\textstyle{1\over 2}} u_n^2 v_{n-1}^2 + u_n v_{n-3} 
\nonumber \\
& & + u_n v_n ( u_n v_{n-2} + 2 u_{n+1} v_{n-1} + u_{n+2} v_n 
+ u_{n+1}^2 v_n v_{n+1}).
\end{eqnarray}

\section{The Mathematica code diffdens.m}

We describe the features, scope and limitations of our program
{\bf diffdens.m}, which is written in {\it Mathematica} syntax \cite{SW}.
The program has its own {\it menu\/} interface with a dozen data files.
Users should have access to {\it Mathematica}.
The code {\bf diffdens.m} and the data files \cite{UGandWHc}
must be in the same directory.

\subsection{The menu interface}

After launching {\it Mathematica}, type
\begin{verbatim}
In[1]:= <<diffdens.m
\end{verbatim}
to read in the code {\bf diffdens.m} and start the program. 
Via its menu interface, the program will prompt the user for answers.

The density is available at the end of the computations. 
To view it in standard {\it Mathematica\/} notation, type \verb|rho|.
To display it in a more elegant subscript notation, 
type \verb|subscriptform[rho]|.

\subsection{Preparing data files}

To test systems that are not in the menu, one should prepare a data file in
the format of the data files that are provided with the code.
Of course, the name for a new data file should not coincide with any name
already listed in the menu, unless one intended to modify those data files.

\vfill
\newpage
\begin{exmp}
{\rm
For the parameterized Toda lattice (\ref{partodalatt}) the data
file reads:
}
\begin{verbatim}
(* Start of data file d_ptoda.m with parameters. *)
(* Toda Lattice with parameters aa and bb *)

u[1][n_]'[t]:= aa*u[2][n-1][t]-u[2][n][t];
u[2][n_]'[t]:= u[2][n][t]*(bb*u[1][n][t]-u[1][n+1][t]);

noeqs = 2;
name = "Toda Lattice (parameterized)";
parameters = {aa,bb};
weightpars = {};

(* The user may give the rank of rho *)  
(* and a name for the output file. *)
(* rhorank = 3; *)
(* myfile = "ptodar3.o"; *)

(* The user can give the weights of u[1] and u[2], *)
(* and of the parameters with weight (if applicable). *)
(* weightu[1] = 1; weightu[2] = 2; weight[aa] = 1; *)

formrho = 0;

(* The user can give the form of rho. *)
(* For example, for the density of rank 3: *)
(* formrho = c[1]*u[1][n][t]^3+c[2]*u[1][n][t]*u[2][n-1][t]+
             c[3]*u[1][n][t]*u[2][n][t]; *)

(* End of data file d_ptoda.m *)
\end{verbatim}
\end{exmp}

\vskip 3pt
\noindent
A brief explanation of the lines in the data file now follows.
\vskip 1pt
\noindent
\verb|u[i][n_]'[t] := ...;|
\vskip 1pt
\noindent
Give the $i^{th}$ equation of the system in {\it Mathematica} notation.
\vskip 1.5pt
\noindent
\verb|noeqs = 2;|
\vskip 1pt
\noindent
Specify the number of equations in the system.
\vskip 1.5pt
\noindent
\verb|name = "Toda Lattice (parameterized)";|
\vskip 1pt
\noindent
Pick a short name for the system under investigation. 
The quotes are essential.
\vskip 1.5pt
\noindent
\verb|parameters = {aa,bb};|
\vskip 1pt
\noindent
Give the list of parameters in the system.
If none, set {\it parameters = \{ \}}.
\vskip 1.5pt
\noindent
\verb|weightpars = {};|
\vskip 1pt
\noindent
Give the list of the parameters that are assumed to have weights.
Note that weighted parameters are {\it not} listed in {\em parameters},
the latter is the list of parameters without weight.
\vskip 1.5pt
\noindent
\verb|(* rhorank = 3; *)|
\vskip 1pt
\noindent
Optional. Give the desired rank of the density, if less interactive 
use of the program is preferred (batch mode). 
\vskip 1.5pt
\noindent
\verb|(* myfile = "ptodar3.o"; *)|
\vskip 1pt
\noindent
Optional. Give a name of the output file, again to bypass interaction 
with the program.
\vskip 1.5pt
\noindent
\verb|(* weightu[1] = 1; weightu[2] = 2; *)|
\vskip 1pt
\noindent
Optional. Specify a choice for {\it some or all} of the weights.
The program then skips the computation of the weights, but
still checks for consistency. Particularly useful if there are several
free weights and non-interactive use is preferred.
\vskip 1.5pt
\noindent
\verb|formrho = 0;|
\vskip 1pt
\noindent
If {\it formrho\/} is set to zero, the program will {\it compute} the 
form of $ \rho_n.$ 
\vskip 1.5pt
\noindent
\begin{verbatim}
formrho = c[1]*u[1][n][t]^3+c[2]*u[1][n][t]*u[2][n-1][t]+
          c[3]*u[1][n][t]*u[2][n][t];
\end{verbatim}
\vskip 1pt
\noindent
Alternatively, one could give a form of $ \rho_n $ (here of rank 3).
The density must be given in expanded form and with coefficients c[i].
If form of $ \rho_n $ is given, the program skips both the computation
of the weights and the form of the density.
Instead, the code uses what is given and computes the coefficients c[i].
This option allows one, for example, to test densities from the literature.
\vskip 1.5pt
\noindent
Anything within \verb|(*| and \verb|*)| is a comment and ignored by
{\it Mathematica}.
\vskip 1pt
\noindent
Once the data file is ready, pick the choice ``\verb|tt) Your System|"
in the menu.

\subsection{Scope and limitations}

Our program can handle systems of first order DDEs that are
polynomial in the dependent variables.
Only one independent variable ($t$) is allowed.
No terms in the DDEs should have coefficients that {\it explicitly} 
depend on $t$ or $n$.
The program only computes polynomial conserved densities in the
dependent variables and their shifts, without explicit dependencies
on $t$ or $n$.

Theoretically, there is no limit on the number of DDEs.
In practice, for large systems, the computations may take a long time
or require a lot of memory.
The computational speed depends primarily on the amount of memory.

By design, the program constructs only densities that are uniform in rank.
The uniform rank assumption for the monomials in $\rho_n$ allows one to
compute {\it independent} conserved densities piece by piece,
without having to split linear combinations of conserved densities.
Due to the superposition principle, a linear combination of conserved
densities of unequal rank is still a conserved density.
This situation arises frequently when parameters with weight are
introduced in the DDEs.

The input systems can have one or more parameters, which are assumed to
be nonzero. If a system has parameters, the program will attempt to
compute the compatibility conditions for these parameters such that
densities (of a given rank) exist.
The assumption that all parameters in the given DDE must
be nonzero is essential.
As a result of setting parameters to zero in a given system of DDEs, 
the weights and the rank of $\rho_n$ might change.

In general, the compatibility conditions for the parameters could be
highly nonlinear, and there is no general algorithm to solve them.
The program automatically generates the compatibility conditions,
and solves them for parameters that occur linearly.
Gr\"obner basis techniques could be used to reduce complicated
nonlinear systems into equivalent, yet simpler, non-linear systems.
For DDEs with parameters and when the linear system for the unknown
coefficients $c_i$ has many equations, the program saves that system 
and its coefficient matrix, etc., in the file {\it worklog.m}.
Independent from the program, the worklog files
can later be analyzed with appropriate {\it Mathematica\/} functions.

The assumption that the DDEs are uniform in rank is not
very restrictive. If the uniform rank condition is violated, the user
can introduce one or more parameters with weights.
This also allows for some flexibility in the form of the densities.
Although built up with terms that are uniform in rank in the 
dependent variables and parameters, the densities do 
no longer have to be uniform in rank with respect to the dependent variables
alone.
Conversely, when the uniform rank condition {\it is} fulfilled, the
introduction of extra parameters (with weights) in the given DDE
leads to a linear combination of conservation laws, not to new ones.

In cases where it is not clear whether or not parameters with weight
should be introduced, one should start with parameters without weight.
If this causes incompatibilities in the assignment of weights (due to
non-uniformity), the program may provide a suggestion. Quite often,
it recommends that one or more parameters be moved from the list
{\it parameters} into the list {\it weightpars} of weighted parameters.

For systems with two or more free weights, the user will
be prompted to enter values for the free weights. If only one
weight is free, the program will automatically compute some choices for
the free weight, and continue with the lowest integer or fractional value.
The program selects this value for the free weight; it is just one choice
out of possibly infinitely many.
If the algorithm fails to determine a suitable value, the user will be
prompted to enter a value for the free weight.

Negative weights are not allowed.
Zero weights are allowed, but at least one of the dependent variables
must have positive weight. The code checks this requirement, and if it
is violated the computations are aborted.
Note that {\it fractional weights} and densities of {\it fractional rank}
are permitted.

Our program is a tool in the search of the first half-dozen
conservation laws. An existence proof (showing that there are
indeed infinitely many conservation laws) must be done analytically.
If our program succeeds finding a large set of independent conservation laws,
there is a good chance that the system of DDEs has infinitely many
conserved densities.
If the number of conserved densities is 3 or less, the DDE may have 
other than polynomial conserved densities, or may not be integrable (in 
the chosen coordinate representation).

\section{Conclusions}

We offer the scientific community a {\it Mathematica} package to carry out
the tedious calculations of conserved densities for systems of nonlinear
DDEs.
The code {\bf diffdens.m}, together with several data and output files,
is available via Internet URL \cite{UGandWHc}.

For lattices with parameters, the code automatically determines the 
compatibility conditions on these parameters so that a sequence of 
polynomial conserved densities exists.

The existence of a large number of conservation laws is an indicator of 
integrability of the lattice.
Therefore, by generating the compatibility conditions, one can analyze 
classes of parameterized DDEs and filter out the candidates for complete 
integrability.

Future generalizations of the algorithm will exploit other symmetries in
the hope to find conserved densities of non-polynomial form. 

\begin{ack}
\"{U}nal G\"{o}kta\c{s} thanks Wolfram Research, Inc. for an internship.
We acknowledge helpful comments from 
Drs. B. Herbst, S. Mikhailov, W.-H. Steeb, Y. Suris, P. Winternitz, and 
R. Yamilov. We also thank G. Erdmann for his help with this project.
\end{ack}



\begin{thebibliography}{30}

\bibitem{EHandKO}
E.G.B. Hohler and K. Olaussen, 
Int. J. Mod. Phys. A 11 (1996) 1831.

\bibitem{ASandRY}
A.B. Shabat and R.I. Yamilov,
Leningrad Math. J. {\rm 2} (1991) 377.

\bibitem{FJH}
F.J. Hickernell,
Stud. Appl. Math. {\rm 69} (1983) 23.

\bibitem{RJLeV}
R.J. LeVeque,
{\rm Numerical methods for conservation laws}
(Birkh\"{a}user Verlag, Basel, 1992).

\bibitem{JMS-S}
J.M. Sanz-Serna,
J. Comput. Phys. {\rm 47} (1982) 199.

\bibitem{DLandPW}
D. Levi and P. Winternitz,
J. Math. Phys. 34 (1993) 3713.

\bibitem{DLandOR}
D. Levi and O. Ragnisco,
Lett. Nuovo Cimento 22 (1978) 691.

\bibitem{ARandBGandKT} 
A. Ramani, B. Grammaticos and K.M. Tamizhmani,
J. Phys. A: Math. Gen. {\rm 25} (1992) L883.

\bibitem{BD} 
B. Deconinck,
Phys. Lett. A {\rm 223} (1996) 45.

\bibitem{ICandRY}
I. Cherdantsev and R. Yamilov,
in: {\rm Symmetries and integrability of difference equations}, 
eds. D. Levi, L. Vinet and P. Winternitz
(American Mathematical Society, Providence, Rhode Island, 1996) 51. 

\bibitem{DLandRY}
D. Levi and R. Yamilov,
J. Math. Phys. {\rm 38} (1997) 6648.

\bibitem{VSandAS}
V.V. Sokolov and A.B. Shabat,
in: Soviet Sci. Rev. Sec. C, Vol. 4
(Harwood Academic Publishers, New York, 1984) 221. 

\bibitem{RY}
R. Yamilov,
in: Proc. 8th int. workshop on nonlinear evolution equations and dynamical
systems (World Scientific Publishing, Singapore, 1993) 423.

\bibitem{UGandWHa}
\"{U}. G\"{o}kta\c{s} and W. Hereman,
J. Symb. Comp. {\rm 24} (1997) 591.

\bibitem{UGandWHb} 
\"{U}. G\"{o}kta\c{s} and W. Hereman. 
The program condens.m is available  
via Internet URL: http://www.mines.edu/fs\_home/whereman/.

\bibitem{UGandWHandGE}
\"{U}. G\"{o}kta\c{s}, W. Hereman and G. Erdmann,
Phys. Lett. A {\rm 236} (1997) 30. 

\bibitem{UGandWHc}
\"{U}. G\"{o}kta\c{s} and W. Hereman.
The program diffdens.m is available from the WWW site in [15].

\bibitem{SW}
S. Wolfram,
The {\it Mathematica\/} book. 3rd Edition
(Wolfram Media, Urbana-Champaign, Illinois \&
Cambridge University Press, London, 1996).

\bibitem{MT}
M. Toda,
{\rm Theory of nonlinear lattices} 
(Springer Verlag, Berlin, 1981).

\bibitem{MAandJLa}
M.J. Ablowitz and J.F. Ladik,
J. Math. Phys. {\rm 17} (1976) 1011.

\bibitem{MAandJLb}
M.J. Ablowitz and J.F. Ladik,
Stud. Appl. Math. {\rm 55} (1976) 213.

\bibitem{MAandBH}
M.J. Ablowitz and B.M. Herbst, 
SIAM J. Appl. Math. {\rm 50} (1990) 339.

\bibitem{XBHandRKB}
X-B. Hu and R.K. Bullough,
J. Phys. A: Math. Gen. {\rm 30} (1997) 3635.

\bibitem{MAandPC}
M.J. Ablowitz and P.A. Clarkson,
Solitons, nonlinear evolution equations and inverse scattering
(Cambridge University Press, Cambridge, U.K., 1991). 

\bibitem{MH} 
M. H\'{e}non,
Phys. Rev. B {\rm 9} (1974) 1921.

\end{thebibliography}
\end{document}